# Deep-Learned Broadband Encoding Stochastic Filters for Computational Spectroscopic Instruments

*Hongya Song, Yaoguang Ma, Yubing Han, Weidong Shen, Wenyi Zhang, Yanghui Li, Xu Liu, Yifan Peng\*, and Xiang Hao\**

H. Song, Dr. Y. Ma, Dr. Y. Han, Prof. W. Shen, W. Zhang, Prof. X. Liu, Dr. X. Hao
State Key Laboratory of Modern Optical Instrumentation,
College of Optical Science and Engineering,
Zhejiang University,
Hangzhou, 310027, China
E-mail: haox@zju.edu.cn

Dr. Y. Li
College of Optical Sicence and Electronic Technology,
China Jiliang University,
Hangzhou, 310018, China

Prof. X. Liu
Collaborative Innovation Center of Extreme Optics,
Shanxi University,
Taiyuan, 030006, China

Dr. Y. Peng
Department of Electrical Engineering,
Stanford University,
Stanford, California 94305, USA
E-mail: evanpeng@stanford.edu

Keywords: optical inverse design; hyperspectral imaging; computational spectroscopy; deep learning

Computational spectroscopic instruments with Broadband Encoding Stochastic (BEST) filters allow the reconstruction of the spectrum at high precision with only a few filters. However, conventional design manners of BEST filters are often heuristic and may fail to fully explore the encoding potential of BEST filters. The Parameter Constrained Spectral Encoder and Decoder (PCSED) – a neural network-based framework is presented for the design of BEST filters in spectroscopic instruments. By incorporating the target spectral response definition and the optical design procedures comprehensively, PCSED links the mathematical optimum and practical limits confined by available fabrication techniques. Benefiting from this, the BEST-filter-based spectral camera present a higher reconstruction accuracy with up to 30 times' enhancement, and a better tolerance on fabrication errors. The generalizability of PCSED is validated in designing metasurface- and interference-thin-film-based BEST filters.

Spectroscopic instruments such as spectrometers and spectral cameras are key drivers of many scientific discoveries in life and material sciences.[1-3] While different technical routes of spectroscopic instruments have been developed and commercially available, a commonly used configuration is that of a filter-based system.[4-6] To pursue a high spectral resolution, these instruments require many filters with sharp but contiguous passing bands, which are practically challenging due to the design and fabrication complexity. Worse still, the narrower passing band is followed by a higher risk of crosstalk and less luminous flux. These limits necessitate the tradeoffs between the spectral resolution and the signal-to-noise ratio (SNR).

To resolve this dilemma, researchers turned to an array of 'broadband encoding stochastic (BEST) filters'. Compared with conventional band-pass ones, BEST filters are advantageous, because they are naturally crosstalk-free, thereby improving the SNR. The production of these filters are also compatible with nearly all new materials, such as quantum dots,[7] plasmonic metasurfaces,[8] photonic crystals,[9] and nanowires.[10] Furthermore, the BEST-filter-based spectroscopic instruments (BEST-SIs), in tandem with advanced computational algorithms,[11] allow the reconstruction of the spectrum with fewer filter channels.[12] As a result, BEST-SIs stand out, for their cost-effectiveness, miniaturization, and portability.[13]

Unfortunately, irrespective of BEST-SI's advantages, an easy-to-follow guideline for designing BEST-SIs is still wanting. The existing attempts always follow a sequential paradigm that includes two individual procedures – first, define the optimal spectral responses mathematically ('target response definition'), and then approach these optima in practice via certain production technique ('optical design'). However, there always lies a contradiction that a mathematically optimal spectral response usually converges to a white-noise-like one, whereas the optical design tends to create sufficiently smooth spectral curves to simplify the production. Regarding this predicament, many BEST-SIs[4, 7-10] are instead built on a so-called 'manufacturing-and-selection' strategy, i.e., a sufficient number of stochastic filters is produced to ensure randomness, and then the BEST filter set is selected via permutation and combination. However, this strategy *de facto* compromises the major advantages of BEST-SI because the number of filters needed as candidates can be orders of magnitude higher than that of the BEST-SI channels. Worse still, the encoding potential of BEST filters can never be fully explored.

Emerging deep learning-based solutions provide an alternative that might be conducive to the BEST-SI design. Aiming at the procedure of target response definition, several works[14, 15] have added artificial priors, for example, Gaussian-shaped or smoothness limit, to avoid converging to the white-noise-like spectral responses. The others[16-19] have enhanced the optical design precision under certain conditions. These methods work to some extent, but still, none of them comprehensively takes both procedures into account. As such, their results are compromised either by the producibility of the designed spectral responses or by the sensitivity to the fabrication error.

Inspired by the emerging data-driven computational imaging approaches,[20-23] we present a novel deep learned framework to comprehensively address both design procedures.

Given a continuous spectrum $S(\lambda)$ in a discrete form

$$\mathbf{s} = [S_1, S_2, \dots, S_N]^T, \tag{1}$$

where *N* is the sampling number, and *T* denotes the transposition operation. When detecting **s** using an *M*-channel BEST-SI, we have

$$\mathbf{y} = W \cdot \mathbf{s}, \tag{2}$$

where $\mathbf{y} = [Y_1, Y_2, \dots, Y_M]^T$ is the intensity vector recorded by all BEST filter channels, and each row of the matrix *W*, $\mathbf{w}_j^T = [w_{j1}, w_{j2}, \dots, w_{jN}]$ is the discretized spectral response of the filter in the *j*-th channel (*j*=1, 2, ⋯, *M*). Note that here in BEST-SI, the filter number *M* can be smaller than *N*. While it represents the sensing matrix in the context of compressive sensing, from deep learning's perspective, the matrix $W$ in Equation 2 acts as a fully connected (FC) layer in an artificial neural network.[24, 25] Each entry $w_{ji}$ in the matrix $W$ corresponds exactly to the connection weighting coefficient from the *i*-th unit in the prior layer to the *j*-th unit in the following layer. Regarding this evidence, the BEST filters form an encoding network (encoder) constructed by a single FC layer (**Figure 1**). By attaching a decoder after it, the whole network, namely, the 'spectral encoder and decoder' (SED) captures and reconstructs the input spectrum. Training the SED grants us access to the value of each $\mathbf{w}_j$, and equivalently, the optimized spectral responses. Accordingly, the training process solves the following optimization problem

$$(\widehat{W}, \hat{\theta}) = \underset{W,\theta}{\mathrm{argmin}} \|\mathbf{s} - \mathcal{D}_\theta(W \cdot \mathbf{s})\|_2^2, \qquad (3)$$

where $\mathcal{D}_\theta(\cdot)$ represents the mapping function of the decoder, $\theta$ corresponds to the network parameters of the decoder, and $\widehat{W}$ represents the optimized spectral responses.

Solving Equation 3 allows a co-design of filter spectral responses $W$ and the decoder parameters $\theta$. On this basis, considering the target response definition and the optical design procedures comprehensively, we aim to add a constraint to $W$ during the training, where the constraint explicitly describes the mapping from the filter structure to its spectral response $\mathbf{w}_j$. We observe that a forward modeling network (FMN)[16, 19] can work perfectly for this task. Notably, it is possible to realize this mapping with other numerical algorithms in specific cases, such as finite difference time domain (FDTD),[26] rigorous coupled-wave analysis (RCWA)[27] or the transfer matrix method.[28] However, FMN is more general and has an advantage in calculation speed (~$10^6$ faster), which is crucial for improving the SED training efficiency (refer to Supplement S1). Besides, as both FMN and SED are artificial neural networks with well-defined gradients, rendering FMN stable and efficient during the SED training. We use a *K*-dimensional vector $\mathbf{p}_j = [p_{1j}, p_{2j}, \dots, p_{Kj}]^T$ to represent $K$ structure parameters (e.g., geometrical sizes of the structures). After training the FMN, the mapping from $\mathbf{p}_j$ to $\mathbf{w}_j$ can be written as

$$\mathbf{w}_j = \mathcal{FM}(\mathbf{p}_j), \tag{4}$$

where $\mathcal{FM}(\cdot)$ denotes the forward mapping carried out by the pre-trained FMN. With $\mathbf{p}_j$ arranged into a $K$-by-$M$ matrix $P = [\mathbf{p}_1, \mathbf{p}_2, ..., \mathbf{p}_M]$, Equation 4 can be written as

$$W^T = \mathcal{FM}(P). \tag{5}$$

By substituting Equation 5 into Equation 3, we have

$$(\hat{P}, \hat{\theta}) = \underset{P,\theta}{\mathrm{argmin}} \|\mathbf{s} - \mathcal{D}_\theta[\mathcal{FM}(P)^T \cdot \mathbf{s}]\|_2^2. \tag{6}$$

We further add regularizations to explicitly limit the desired structure parameter range, which is beneficial for making the design robust to fabrication errors. Hence, Equation 6 becomes

$$(\hat{P}, \hat{\theta}) = \underset{P,\theta}{\mathrm{argmin}} \|\mathbf{s} - \mathcal{D}_\theta[\mathcal{FM}(P)^T \cdot \mathbf{s}]\|_2^2 + R(P), \tag{7}$$

where $R(P)$ is the regularization term. The above procedure gives birth to a parameter constraining the spectral encoder and decoder (PCSED) neural network (Figure 1). This framework allows the structure parameters of BEST filters to be directly trained, which is impossible in the sequential design paradigm. Therefore, PCSED allows a full exploration of the optical designing flexibility. Furthermore, the PCSED leverages back-propagation through an FMN to obtain the learned structure parameters. Because the forward modeling mapping is injective (where there is a one-to-one mapping between one physical system and its corresponding response), PCSED is better convergent than the sequential inverse design methods[16] for the same task.

To support our claim, we applied the PCSED to the design of BEST filter-based spectroscopic instruments. In our first demonstration, we designed a metasurface BEST filter-based spectral camera. We set the number of BEST filters as 4, an extension of a Bayer filter unit[29] over the imaging sensor. In this way, the PCSED simulates the working stream of a metasurface spectral camera – the four adjacent metasurface BEST filters construct a macro-pixel and encodes the spectrum of the incident light, and the detected intensity data is decoded by the decoder to reconstruct a spectrum. We used both CAVE[30] and ICVL[31] datasets for training the PCSED, by which the spectral range of interest is 400-700 nm. The training and the test sets respectively include $10^6$ and $10^5$ spectral samples, which are randomly selected from the pixels of CAVE and ICVL hyperspectral images. The spectral images in the original CAVE and ICVL datasets have 31 spectral channels, corresponding to a spectral step size of 10 nm. Because the sampling number of each BEST filter's response must be the same as the spectral channel number of the dataset, we interpolated them into 151 channels, corresponding to a spectral step size of 2 nm. The interpolation is essential, because only in this way, the spectral responses of the deeply learned filters can be finely defined, and the network simulates the physical world accurately. Notably, the spectral resolution of the whole network is not improved, because the interpolation does not introduce additional information, thus the spectral step of the output data does not represent the spectral resolution.

As shown in **Figure 2**(a), we synthetically built the metasurface BEST filters with a basic structure of silicon nano-bricks on a 2-layer substrate, which is constructed by coating a $Si_3N_4$ thin layer on a $SiO_2$ block. The structure parameters that potentially influence the spectral responses involve the edge length (L), the height (H) of the nano-bricks, the thickness (T) of the $Si_3N_4$ layer, and the unit period (D). Each parameter varies in a range of 100–200 nm (L), 50–200 nm (H), 200–400 nm (T), and 300–400 nm (D), respectively. Tuning the structure parameters enables us to obtain diverse spectral responses.[32, 33] (For the training details of FMN and PCSED, see Supplement S2.1 and S2.2.)

For comparison, we also applied a sequential design paradigm for the same task. In detail, we built a SED and added an $l_2$-norm-based smoothness regularization to the single-FC-layered encoder (method in Ref. [15]). After training the SED, we fed the trained target spectral responses into an inverse design network to derive the structures of metasurface BEST filters (method in Ref. [19]). For brevity, we denote this method 'SED-inv'. (For details of SED-inv, see Supplement S2.3.)

After running both PCSED and SED-inv, we fed the output structure parameters into the FDTD simulation to generate the spectral responses of the designed BEST filters. We further added the random structure variance within the range of ±3 nm to evaluate the possible fabrication error. By taking the target, the designed BEST filters, and those added with fabrication errors as the encoder, the spectrum reconstructed by the decoder and the corresponding reconstruction error level are assessed using the mean square error (MSE).

Figure 2(b-c) present the spectral responses of all four BEST filters generated by PCSED [Figure 2(b)] and SED-inv [Figure 2(c)], respectively. While two curves from PCSED roughly match on the shape, the spectral responses (dashed orange) designed by SED-inv present strong fluctuations around the target responses (solid orange). Applying PCSED can effectively reduce the MSE between the target and the design responses by a factor of 2-10. Not surprisingly, lower MSE in design induces higher precision of spectral reconstruction. **Table 1** highlights the spectral reconstruction MSE of the metasurface spectral camera designed by both methods. In particular, the MSE values in the 'Target' row are extremely small ($<10^{-3}$), suggesting that both methods are applicable for training the BEST filters and the decoder. However, the producibility of the trained BEST filters differs. When taking the designed BEST filters as the encoder, the reconstruction MSE of the PCSED still maintains at a low level ($\sim 10^{-3}$), while that of the SED-inv is approximately 30 times higher ('Designed' row). On the other hand, the MSE contributed by the fabrication error is minute. For the PCSED, the fabrication error only increases the MSE by ~10%, while for the SED-inv, the MSE is even smaller after we introduced the artificial fabrication error ('W. fab. error' row). As the mismatch between the target and the designed spectral responses dominates the reconstruction MSE, employing PCSED instead of SED-inv becomes critical.

To visualize the spectral reconstruction results, we simulated the imaging process of the designed camera and analyzed the reconstructed spectrums of the hyperspectral images. Figure 2(d-h) displays the results when we used the designed BEST filters [dashed curves in Figure 2(b) and (c)] as the encoder. For the results using the BEST filters with artificial fabrication errors, refer to the supplement S2.4. For all 7 images randomly selected from CAVE and ICVL datasets [Figure 2(d)], while the SED-inv fairly reconstructs the spectrum [Figure 2(f)], the PCSED strikingly increases the overall peak signal-to-noise ratio (PSNR) for more than 10 dB [Figure 2(e)]. In contrast to Figure 2(e) in which the purple color is almost uniformly distributed, Figure 2(f) includes more components with larger MSEs (highlighted with bright colors). Even for the positions where the MSEs of both methods are relatively low (marked by red and green patches), our method is superior in that it offers a higher spectral reconstruction precision [Figure 2(g-h)].

Remarkably, PCSED is not limited to optimizing metasurfaces. In principle, it applies to any production techniques whose spectral response is theoretically predictable via a well-defined mathematical model. In our second example, we aimed to design the thin-film interference BEST filters. We assumed that these filters have 5 layers of $TiO_2$ and 5 layers of $SiO_2$ alternatively coated on a glass substrate. In PCSED, the FMN enabled us to conveniently limit the thickness of each layer to a range of 100-300 nm, approximately a quarter wavelength if the light is visible. Such a scale is also desirable in avoiding super-thin layers that are challenging for fabrication.[34] Similarly, we have also developed the SED-inv networks and then compared the results with those from our PCSED method (Supplement S3.1–S3.3). Intuitively, the designed spectral responses from both methods roughly match the target ones ('designed' column in Table S3), but still, PCSED is significantly better. In addition to a ~50% smaller MSE of the BEST filters' spectral responses, PCSED also exhibits a better tolerance to fabrication errors. Specifically, the BEST filters designed by PCSED decrease the MSE by a factor of ~7.5 when taking those from SED-inv for reference (see Supplement S3.4).

In summary, we have proposed a novel deep-learned co-design of the BEST filters and the decoder in computational spectroscopic instruments. By representing the array of BEST filters as an FC layer of an artificial neural network, it is viable to optimize the BEST filters' spectral responses by training the entire network. Furthermore, using the FMN to conduct the mapping from structure parameters to the spectral responses enables us to improve the design convergence. In the meanwhile, the design procedure is simplified by evading the further inverse design, irrespective of the filter production methods. Simulation results prove that our design paradigm bridges the gap between the target response definition and the optical design procedures, thereby enabling more flexibility in design space. Although the current output of the PCSED involves only the reconstructed spectrum where we foresee an urgent application demand, we envision the proposed method applies to alternative visual tasks such as spectral recognition or classification.

**Acknowledgements**
The authors gratefully acknowledge the financial support from the National Key R&D Program of China (2018YFA0701400), the Major Research Plan of the National Natural Science Foundation of China (92050115), the Fundamental Research Funds for the Central Universities (2019QNA5006), ZJU-Sunny Photonics Innovation Center (2019-01), and Zhejiang Lab (2020MC0AE01). The authors thank Mr. Chang Wang and Mr. Mingwei Tang for the instructions on the FDTD simulation.

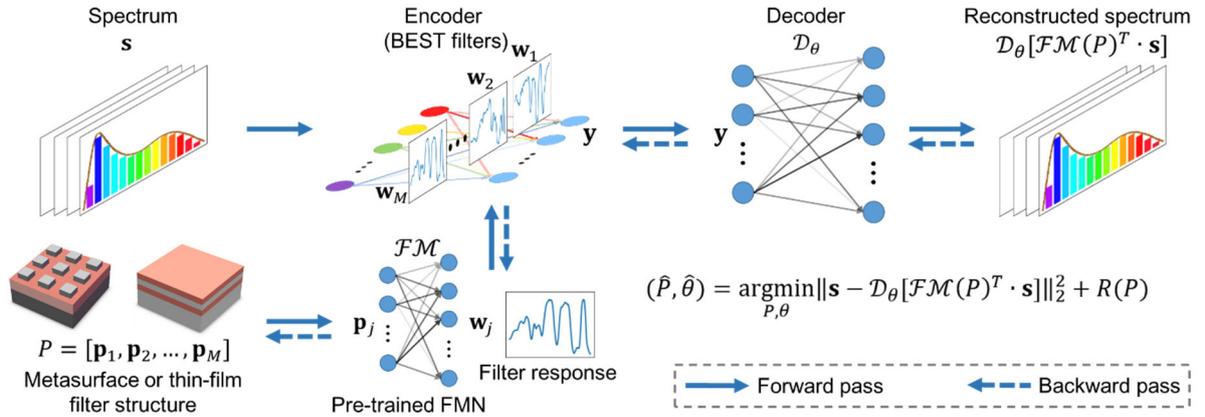

**Figure 1.** Schematic of our parameter constrained spectral encoder and decoder (PCSED) design framework. The random optical filters (ROFs) act as an encoding neural network (encoder), whose connection weights (corresponding to the ROFs' responses) are constrained by the filters' structure parameters through a pre-trained forward modeling network (FMN). During each training epoch, a batch of spectrums is fed into the encoder and the decoder gives the corresponding output, such as the reconstructed spectrums. The loss function, such as the mean square error (MSE) for the reconstructed spectrum is evaluated and the errors are back-propagated to the structure parameters (e.g., geometrical parameters of metasurface or thin-film structures). Under this framework, the ROFs and the decoder are jointly designed.

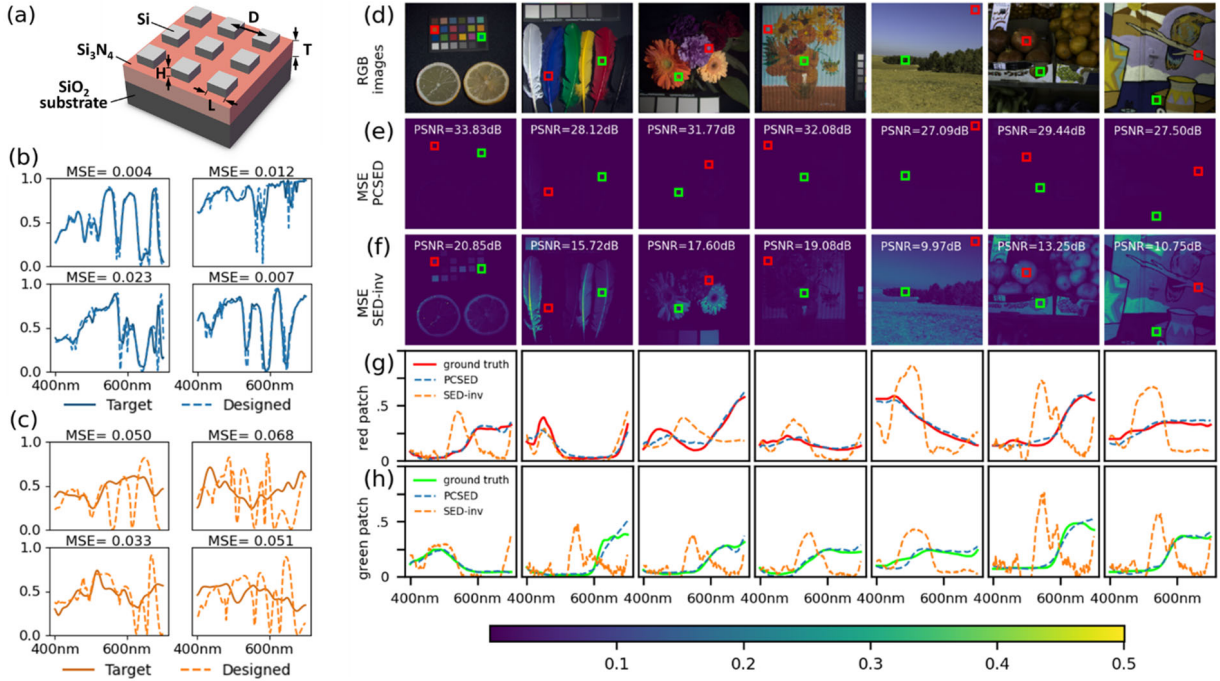

**Figure 2.** Schematic and synthetic performance of the metasurface BEST-filter-based hyperspectral camera. (a) Schematic of BEST filter's nanostructures. The structure parameters include the edge length (L) and height (H) of nano-bricks, the thickness (T) of the $Si_3N_4$ layer, and the period (D) of the repeating units. (b, c) Spectral responses of the target (solid curves) and designed (dashed curves) BEST filters generated by (b) PCSED and (c) SED-inv. (d-h) Synthetic hyperspectral reconstruction by using the designed cameras. For each row, there are (d) hyperspectral images in RGB form, chosen from CAVE (first four) and ICVL (last three); (e, f) the pixel-wise spectral reconstruction MSE of each hyperspectral image, the value on top of each figure stands for the peak signal to noise ratio (PSNR) of the whole image; (g, h) the reconstructed spectrum from the red and green patches in (d).

**Table 1.** Spectral reconstruction MSE of the metasurface spectral camera designed by PCSED and SED-inv [a]

|  | PCSED | SED-inv |
| --- | --- | --- |
| Target | 0.00015 | 0.00009 |
| Designed | 0.0011 | 0.0358 |
| W. fab. error | 0.0013 | 0.0342 |

[a]MSEs are averaged from that of samples within the corresponding sets.

Computational spectroscopic instruments have great potential on integration, cost-effectiveness and portability. However, the optimization and the inverse design of suitable filters is still challenging. Here a novel deep-learning-based design method is proposed. It offers an end-to-end optimization pipeline for both the filter and software design, and improves the design accuracy significantly.

**Keyword** optical inverse design

Hongya Song, Yaoguang Ma, Yubing Han, Weidong Shen, Wenyi Zhang, Yanghui Li, Xu Liu, Yifan Peng*, and Xiang Hao*

**Deep-Learned Broadband Encoding Stochastic Filters for Computational Spectroscopic Instruments**

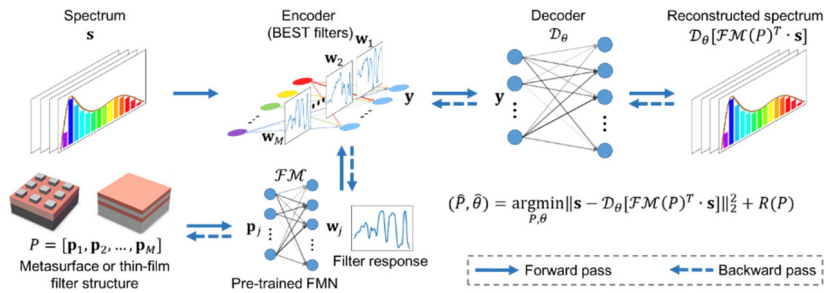

Supporting Information

**Deep-Learned Broadband Encoding Stochastic Filters for Computational Spectroscopic Instruments**

*Hongya Song, Yaoguang Ma, Yubing Han, Weidong Shen, Wenyi Zhang, Yanghui Li, Xu Liu, Yifan Peng\*, and Xiang Hao\**

## S1 Run-time comparison between FMN and FDTD

Training PCSED usually requires hundreds of epochs with the forward modeling being carried out during each epoch. As such, fast forward modeling is highly appreciated. For comparison, we ran the forward modeling network (FMN) and the finite difference time domain (FDTD) method on the full training and test dataset to assess the average run-time for each sample. The dataset is the one used for training the FMN (i.e., 6561 samples in total). The FDTD simulation is conducted using the commercial software 'Lumerical FDTD Solutions'.

First, we ran each method on the same platform (CPU: Intel(R) Core(TM) i3-6100, 3.70 GHz, 2 Cores; RAM: 12GB). The result is shown in the first row of Table S1. The FMN is ~$2.7 \times 10^6$ times faster than that of FDTD. After that, regarding the fact that training PCSED is usually implemented on a GPU platform, we re-ran the FMN on an Nvidia GeForce RTX2080Ti platform for additional comparison. Note that the available version of FDTD simulation software is not compatible with GPU computing. Thus, we re-ran the FDTD simulation on a computational server platform (CPU: Intel(R) Xeon(R) Gold 5115, 2.40 GHz, 2CPUs, 20 Cores; RAM: 128GB). The result is shown in the second row of Table S1. With the GPU implementation, the FMN is about $2.1 \times 10^8$ times faster than that of FDTD. As such, it is clear that the processing of FMN is faster than that of FDTD method. Moreover, the FMN is less expensive on GPU-CPU data transmission when embedded onto training PCSED.

**Table S1 Average computation time for one sample using both FMN and FDTD method**

| Platform \ Method | FMN | FDTD |
|---|---|---|
| Same (Core i3 CPU) | 47.4 us | 127.3 s |
| Different (Nvidia GPU or Xeon CPU) | 1.01 us | 21.6 s |

## S2 Additional details of learning a metasurface hyperspectral camera

**S2.1 FMN.** The architecture of FMN is denoted as 4-FC-200-BN-LR-FC-800-BN-LR-FC-800-DO-BN-LR-FC-800-DO-BN-LR-FC-800-DO-BN-LR-FC-151-DO-Sigmoid, where each number indicates the number of units in the corresponding layer. Abbreviations: FC: a fully connected layer, BN: a batch normalization layer, DO: a dropout layer with a dropout ratio of 0.1, LR: leaky ReLU units, Sigmoid: sigmoid units. The 4 input units correspond to the 4 geometrical parameters (L, H, T, D, see the main text) of the metasurface structure. The 151 output units stand for the 151 output spectral channels (400 nm to 700 nm with 2 nm interval). The loss function is mean square error (MSE) loss. The spectral responses were calculated using the commercial software Lumerical FDTD Solutions to build up 6561 samples. We randomly selected 80% (5248) of the samples for training and the rest 20% (1313) for test. We test the FMN every 10 epochs to monitor overfitting and we do not change any hyperparameters during a training process. Therefore, the test set is still suitable for evaluating the generality. The same strategy is carried out during the training of all the other networks. Figure S1 shows the training curves and the performance. The trained FMN has an MSE of ~0.00398 after 2000 epochs of training. The training process takes 44s on a Linux platform (CPU: Intel(R) Core(TM) i7-9800X, 3.80 GHz, 8 Cores; RAM: 125GB; GPU: Nvidia GeForce RTX2080Ti).

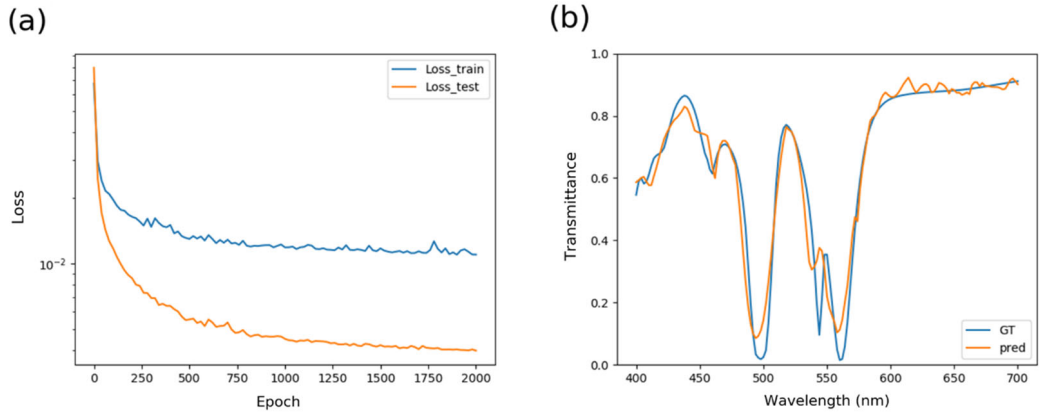

Figure S1. Performance of FMN training. (a) Training curve of FMN. The blue curve is the training loss, while the orange one is the test loss. Note that the test loss is smaller than the training loss. This is because the dropout is turned off when testing, which leads to a better fitness. (b) Prediction of a filter's spectral response, randomly chosen from the test set. The blue curve is the ground truth, while the orange one is the output of FMN.

**S2.2 PCSED.** The architecture of PCSED is denoted as 151-FMFC-4-LR-FC-500-LR-FC-500-LR-FC-151-LR, where the 'FMFC' indicates a fully connected layer linked by an FMN. In this architecture, the FMFC layer connects between 151 input units and 4 hidden units, acting like 4 BEST filters. For each BEST filter $j$, the connection weights $\{w_{j1}, w_{j2}, ..., w_{j151}\}$ are generated by feeding the 4 structure parameters $\{p_{1j}, p_{2j}, p_{3j}, p_{4j}\}$ into the FMN. To avoid the trained structure parameters from exceeding the desired range, a regularization term is added to the cost function, which is denoted as

$$\beta_r \frac{1}{M \cdot P} \sum_{j=1}^{M} \sum_{k=1}^{P} \max \begin{cases} \frac{(lb_j + \delta) - p_{kj}}{\delta} \\ 0 \\ \frac{p_{kj} - (ub_j - \delta)}{\delta} \end{cases} \tag{S1}$$

where $M$ is the number of BEST filters ($M = 4$), $P$ is the number of the structure parameters ($P = 4$), $lb_j$ stands for the lower bound of the $j$-th structure parameters, and $ub_j$ is for the upper bound. In our case, the lower and upper bounds are set to be the geometrical size range of the generated training BEST filters, which is 100 to 200 nm ($L$), 50 to 200 nm ($H$), 200 to 400 nm ($T$), and 300 to 400 nm ($D$). $\delta$ is a small number to control the slope (in our case, $\delta = 0.01$). $\beta_r$ is a hyperparameter that controls the proportion of this regularization term in the cost function, which can be adjusted before training. When the structure parameters are between $t_{min} + \delta$ and $t_{max} - \delta$, the regularization term is 0. When the structure parameters exceed this range, the regularization term grows. In this way, the structure parameters can be limited between the lower and upper bounds, which helps satisfy a practical fabrication demand.

Both the input and the output unit number of PCSED are 151, which means the PCSED takes the discretized spectrum as input to export the reconstructed spectrum at the same resolution. The loss function calculates the MSE between the output and the input spectrum. The training dataset includes $10^6$ samples and the test dataset includes $10^5$ samples, respectively. Samples in both sets are spectral curves randomly chosen from the pixels of the hyperspectral images in CAVE and ICVL dataset (half from CAVE and half from ICVL). The training curve is shown in Figure S2. The trained PCSED has an MSE of ~0.00015 after 500 epochs of training. The training process takes 2,673 s (44'33") on a Linux platform (CPU: Intel(R) Core(TM) i7-9800X, 3.80 GHz, 8 Cores; RAM: 125GB; GPU: Nvidia GeForce RTX2080Ti).

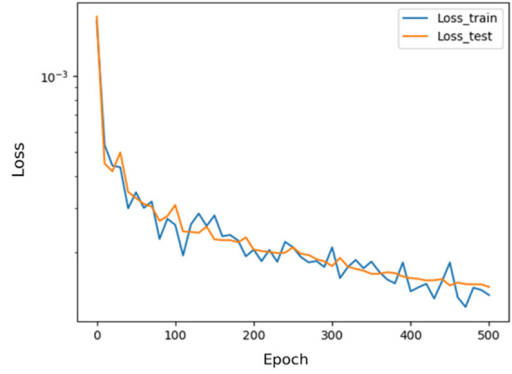

Figure S2. Training curve of PCSED. The blue curve is the training loss while the orange one is the test loss.

**S2.3 SED-inv.** The architecture of SED is identical to PCSED except for removing the FMN constraint from the single-FC-layer encoder (FMFC layer). The architecture of SED is denoted as 151-FC-4-LR-FC-500-LR-FC-500-LR-FC-151-LR. A regularization term is also added to the cost function. Let $\{w_{j,i}\}$ represent the weights of the encoding FC layer that connects the $i$-th unit in the input layer and the $j$-th unit in the next layer, the regularization is constructed as

$$\beta_r \frac{1}{M \cdot N} \sum_{i=1}^{N} \sum_{j=1}^{M} \max \begin{cases} \frac{(0+\delta) - w_{j,i}}{\delta} \\ 0 \\ \frac{w_{j,i} - (1-\delta)}{\delta} \end{cases} + \beta_s \sqrt{\sum_{i=1}^{N-1} \sum_{j=1}^{M} (w_{j,i} - w_{j,i+1})^2} \quad (S2)$$

where $M$ is the number of BEST filters ($M = 4$), $N$ is the number of spectral channels ($N = 151$), and $\delta$ is a small number (In this case, $\delta = 0.01$). $\beta_r$ and $\beta_s$ are hyperparameters that control the proportion of each regularization term in the cost function, which can be adjusted before training. The first term indicates the range limit, keeping the optimized spectral transmittance responses limited between 0 and 1. The second term is an $l_2$-norm-based regularization, controlling the smoothness of each BEST filter's spectral response curve.

The dataset for training and test the SED is the same as that of the PCSED, and the loss function is also MSE loss. The training curve is shown in Figure S3. The trained SED has an MSE of ~0.00009 after 500 epochs of training. The training process takes 2,279 s (37'59") on a Linux platform (CPU: Intel(R) Core(TM) i7-9800X, 3.80 GHz, 8 Cores; RAM: 125GB; GPU: Nvidia GeForce RTX2080Ti).

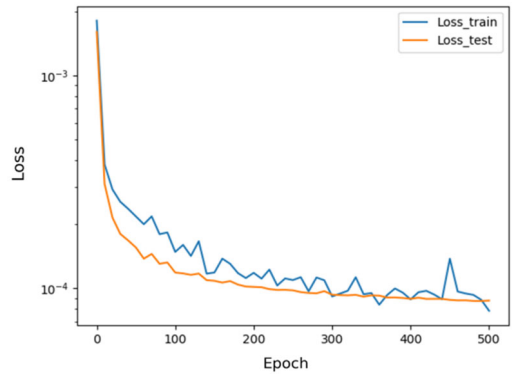

Figure S3. Training curve of SED. The blue curve indicates the training loss and the orange curve indicates the test loss.

After training the SED, the BEST filters are designed using a deep-learning-based inverse design method presented in Ref. [19] of the main text. It utilizes a tandem neural network architecture, which is to build an inverse design network (IDN) followed by a pre-trained FMN to keep the training process converge. The architecture of the IDN is denoted as 151-FC-2000-BN-LR-FC-2000-BN-LR-FC-800-BN-LR-FC-800-BN-LR-FC-100-BN-LR-

FC-4-Sigmoid. The dataset for training and test the inverse design network is the same as the dataset of the FMN. The loss function is MSE loss. The training curve and performance are shown in Figure S4. The trained IDN has an MSE of about 0.01735. Compared with the FMN, the IDN has a more complicated architecture. However, the inverse design accuracy is ~4 times lower (the forward modeling MSE of FMN is 0.00398). It confirms the fact that the inverse design performs less convergence than the forward modeling. Therefore, the SED-inv method requires a more complicated inverse design process than the PCSED to achieve the same design accuracy.

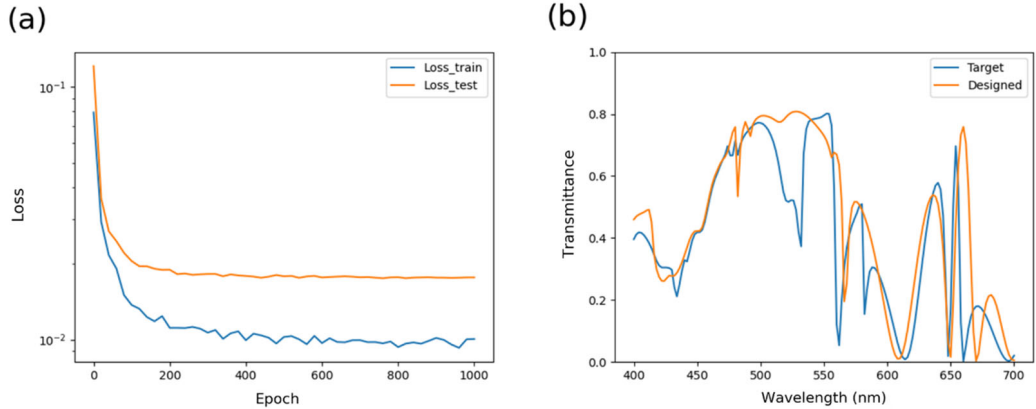

Figure S4. Performance of SED-inv. (a) Training curve of inverse design network. The blue curve is the training loss, and the orange curve is the test loss curve. (b) Target (blue) and designed (orange) spectral response using inverse design network. The target spectral response is randomly chosen from the test set, and the designed spectral response is acquired by an FDTD simulation of the inverse designed structure.

**S2.4 Performance comparison between PCSED and SED-inv.** Figure S5 shows the spectral responses of all four BEST filters generated by PCSED [Figure S5 (a)] and SED-inv [Figure S5 (b)]. We not only compare the BEST filter designing accuracy of the two methods but also synthetically add ±3 nm maximum random fabrication errors to analyze the fabrication error tolerance of both designs. For each filter, we use the FDTD simulation software to build a large metasurface simulation unit, which covers 400 nano-brick units. We add random geometrical errors to each structure parameter, including the thickness of the $Si_3N_4$ layer ($T$) the height ($H$) of each nano-brick, the edge length ($L$) in $x$, $y$ directions and the location ($D$) shift in $x$, $y$ directions of each nano-brick. We also add +3 nm and –3 nm geometrical errors to the BEST filter structures for a large-error situation analysis. In Figure S5, the spectral responses added with fabrication errors is denoted as dashed curves. In Figure S5 (a), the blue solid curves fit the black ones in shape, while in Figure S5 (b), the fitness drops dramatically. Moreover, in Figure S5 (a), the orange dashed curves also fits the black ones well. It reveals that the PCSED method offers both a high design accuracy and an acceptable tolerance on fabrication errors. Note that in Figure S5 (b), the 'Fab. MSE' of some BEST filters are smaller than the 'Design MSE'. This is because the design error when using the SED-inv method is so large that random fabrication error may push the spectral response slightly closer to the target response. However, the overall spectral reconstruction accuracy is still dominated by the mismatch between the target and the designed spectral responses (See Table 1 in the main text and Figure S7). Figure S6 and S7 are synthetic hyperspectral reconstruction by taking the target BEST filters and the designed BEST filters added with fabrication errors as the encoder. The reconstruction MSE level and the overall PSNR are consistent with the data in the main text Table 1. Specifically, even using the BEST filters added with fabrication errors, the PCSED can also reconstruct the shape of the spectrums (Figure S7), which suggests our PCSED method is suitable for practical BEST filter-SI designing.

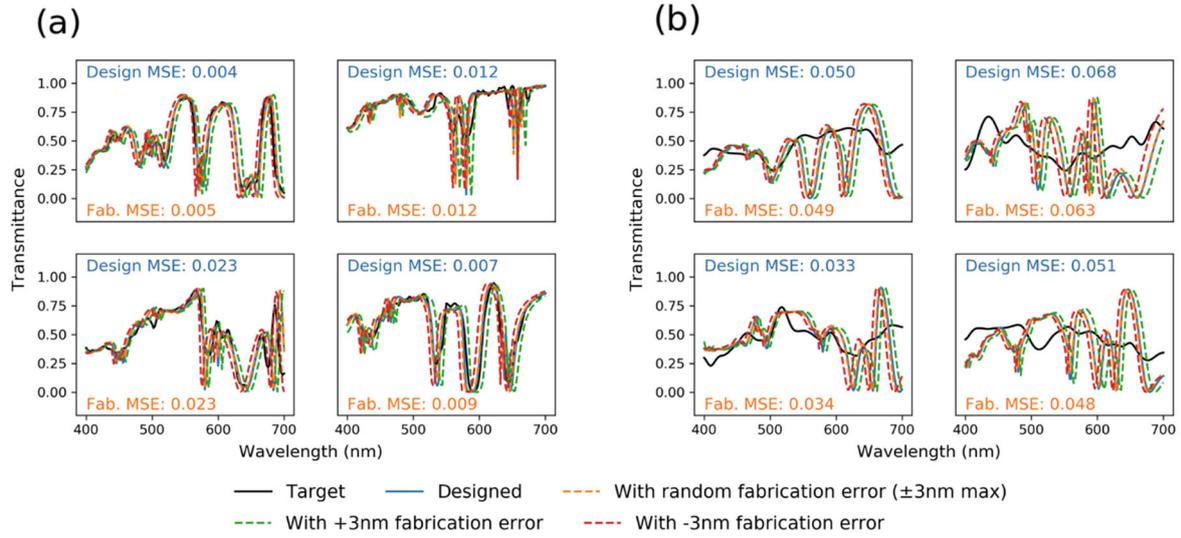

Figure S5. Spectral responses of the target (black solid), designed (blue solid), and synthetically fabricated (dashed) metasurface BEST filters using (a) PCSED and (b) SED-inv methods. The orange dashed curves are responses that simulated by adding ±3 nm maximum random fabrication errors to the designed structure parameters, the green and the red dashed curves are with +3 nm and –3 nm errors, respectively. The 'Design MSE' labeled at the top of each graph represents the MSE between the designed and the target spectral response for each BEST filter and the 'Fab. MSE' labeled at the bottom represents the MSE between the synthetically fabricated and the target spectral response.

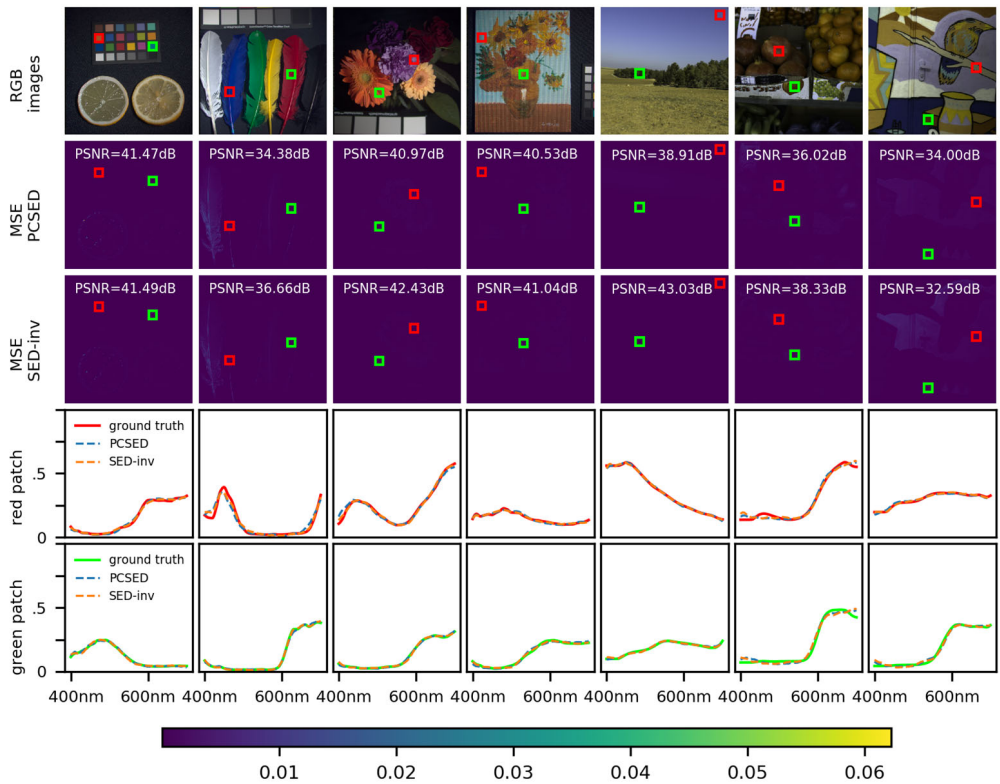

Figure S6. Hyperspectral image reconstruction using the target BEST filters by PCSED and SED-inv method. As there is no design error, both networks reconstruct the hyperspectral images with high accuracy.

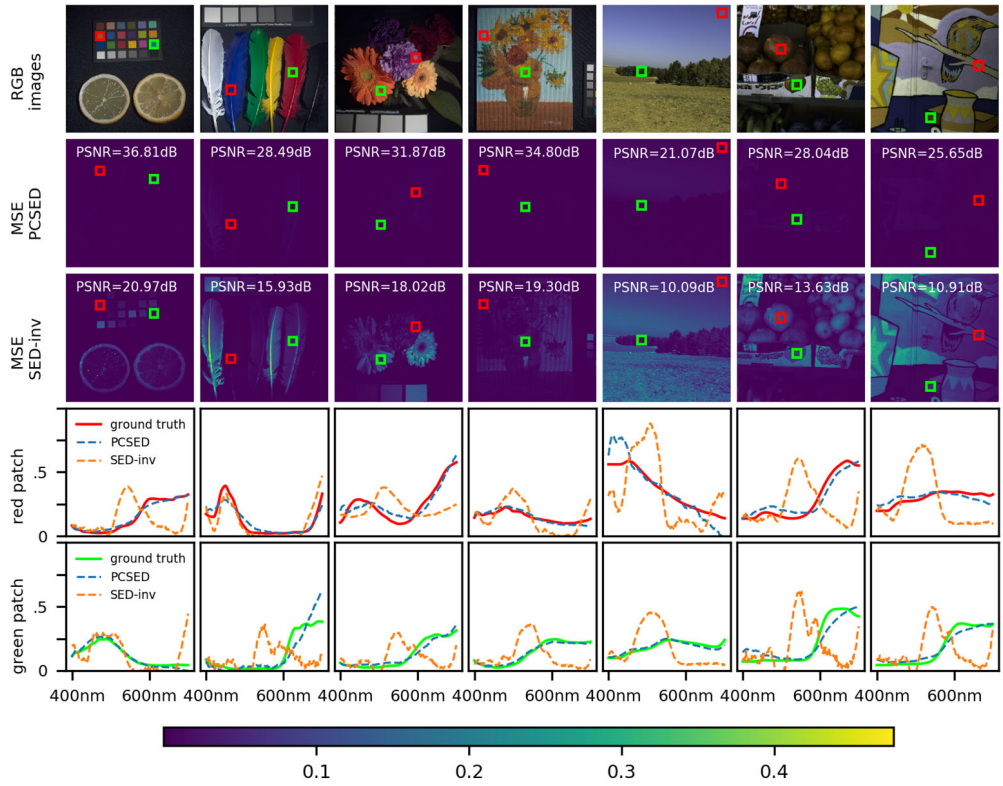

Figure S7. Hyperspectral image reconstruction using the BEST filters with a ±3 nm maximum random fabrication error by PCSED and SED-inv method. With design error and the fabrication error, the performance of the PCSED is significantly better than that of the SED-inv.

# S3 Additional details of learning a thin-film hyperspectral camera

**S3.1 FMN.** The dataset is generated by a planar multilayer thin-film simulation program.[1] The training dataset includes 500,000 samples, and the test dataset includes 50,000 samples, respectively. The architecture of FMN is denoted as 10-FC-200-BN-LR-FC-800-BN-LR-FC-800-BN-LR-FC-800-BN-LR-FC-800-BN-LR-FC-151-Sigmoid. The input units represent the 10 thicknesses of $SiO_2/TiO_2$ layers, while the 151 output units are for the spectral channels (400 nm to 700 nm with 2 nm interval). The loss function is set to be the MSE loss. Figure S8 shows the training curve and performance. The trained FMN has an MSE of ~0.0006 after 2000 epochs of training. The training process takes 8,153 s (2h15'53") on a Linux platform (CPU: Intel(R) Core(TM) i7-9800X, 3.80 GHz, 8 Cores; RAM: 125GB; GPU: Nvidia GeForce RTX2080Ti).

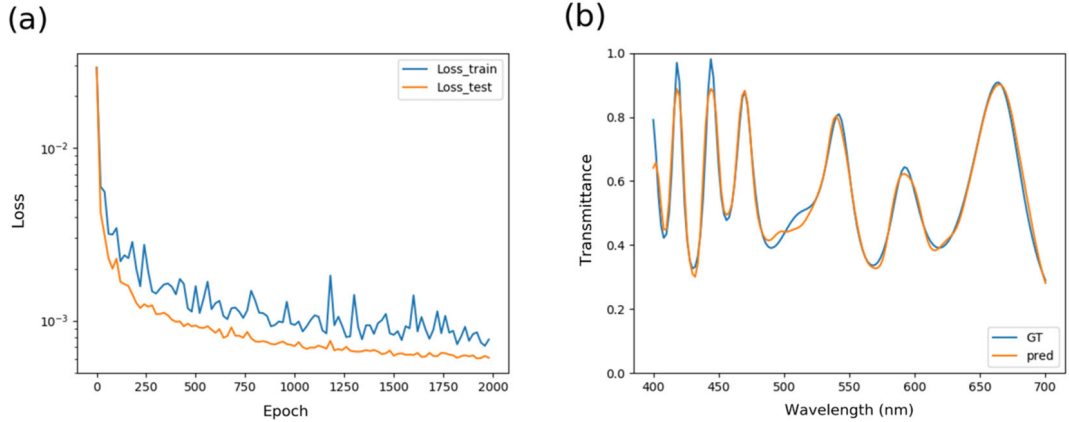

Figure S8. Performance of FMN training. (a) Training curve of FMN. The blue curve is the training loss, and the orange curve is the test loss. (b) Prediction of a filter's spectral response, randomly chosen from the test set. The blue curve indicates the ground truth, and the orange curve is for the output of FMN.

**S3.2 PCSED.** The architecture of PCSED is denoted as 151-FMFC-4-LR-FC-500-LR-FC-500-LR-FC-151-LR. To avoid the trained thicknesses from violating the thickness constraint (100 nm to 300 nm in this example), the regularization term (equation S1) is added to the cost function, in which the lbi is 100 nm and the ubi is 300 nm for all i=1, 2, 3, 4. The loss function is MSE loss. The training and the test spectrum datasets are the same as the ones used in the metasurface hyperspectral camera. The training curve is shown in Figure S9. The trained PCSED has an MSE of ~0.00008 after 500 epochs of training. The training process takes 1,600s (26'40") on a Linux platform (CPU: Intel(R) Core(TM) i7-9800X, 3.80 GHz, 8 Cores; RAM: 125GB; GPU: Nvidia GeForce RTX2080Ti).

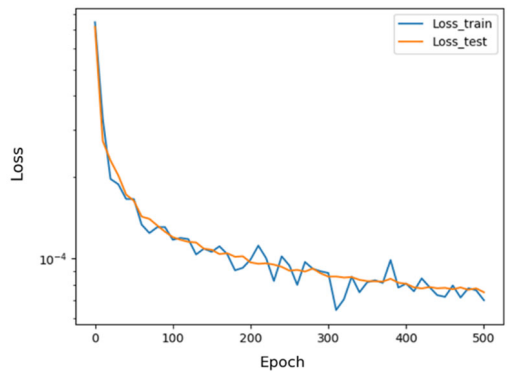

Figure S9. Training curve of PCSED. The blue and orange curves indicate the training loss and the test loss, respectively.

We employed the same SED architecture and the same target filter responses in that of the metasurface hyperspectral camera but redesigned the BEST filters using thin-film structure. When carrying out the inverse design in SED-inv, there is a trade-off between the inverse design accuracy and the fabrication robustness. The target spectral responses that learned by the SED are usually 'smooth' because of the l2-norm-based regularization. Besides, due to the interference when the light is reflected between the thin-film layers, thicker layers always result in larger ripples on the spectral response curves.[2] Therefore, thinner film layers are preferred to make the inverse design network easier to converge and fit the desired smooth responses more accurately. However, very thin layers (less than 20 nm) are also avoided in the design, because there are many practical difficulties in very-thin-film formation.[3] This makes the designed BEST filters less robust to the fabrication perturbations. Therefore, there lies a trade-off in the SED-inv method – it is difficult to realize both accurate and robust design results. For a comprehensive comparison, we trained two inverse design networks – one with a thickness constraint of 0 to 150 nm (IDN1) to achieve a relatively high design accuracy and another with a thickness constraint of 100 nm to 300 nm (IDN2) to achieve relatively high fabrication error tolerance. The architectures of both IDNs are the same, which is denoted as 151-FC-2000-BN-LR-FC-2000-BN-LR-FC-800-BN-LR-FC-800-BN-LR-FC-100-BN-LR-FC-10-Sigmoid.

The dataset for training the IDN1 is also generated by the numerical simulation program[1] with randomly varying thicknesses in between 0 to 150 nm, and the dataset for training the IDN2 is the same as that in S3.1 used for training the FMN. Both datasets have 500,000 training samples and 50,000 test samples. The loss function is MSE loss. The training curve and performance are shown in Figure S10. The trained IDN1 has an MSE of ~0.00146, and the IDN2 has an MSE of ~0.01352.

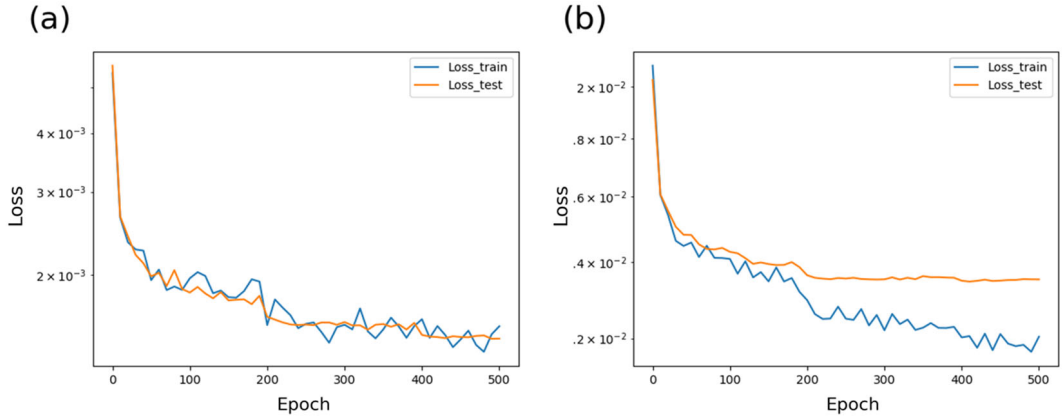

Figure S10. Training curves of (a) IDN1 and (b) IDN2. The blue and the orange curves are the loss curves for the training and the test, respectively.

**S3.4 Performance comparison between PCSED and SED-inv.** Figure S11 gives the spectral responses of all four BEST filters generated by PCSED [Figure S11 (a)], SED-inv with IDN1 [Figure S11 (b, d)], and SED-inv with IDN2 [Figure S11 (c)], respectively. In Figure S11 (a) and (b), the designed spectral responses (blue solid curves) can roughly fit the target ones (black solid curves), and the design accuracy (shown as the 'Designed MSE' labeled at the top) of PCSED is still ~2 times higher than that of the SED-inv. Moreover, the spectral responses of BEST filters with fabrication errors (orange, red, and green dashed curves) are closer to the designed and target ones using PCSED than the SED-inv [Figure S11 (a, b and d)]. With a ±3 nm maximum random thickness error added to the design process (the orange dashed curves), our PCSED strategy shows 1.3-2.3 times higher accuracy (the 'Manu. MSE' marked at the bottom) than the SED-inv method. This means our PCSED method shows not only higher design accuracy, but also larger fabrication error tolerance. In Figure S11 (c), the inverse design accuracy drops over 5 times than Figure S11 (b). This comparison confirms the trade-off between the design accuracy and the robustness – a forced thickness constraint of 100–300 nm leads to a high design error, while a thickness range of 0–150 nm can afford acceptable design accuracy but is more sensitive to the fabrication errors.

To quantitatively evaluate the spectral reconstruction error induced by the fabrication errors, we replaced the encoders in PCSED and SED-inv methods with the designed (blue solid curves) and BEST filters added with random fabrication errors (orange dashed curves), and re-ran the networks without secondary training. Notably, for SED-inv method, we chose the BEST filters designed by IDN1, where the design accuracy is higher than the IDN2. The spectral reconstruction MSE is shown in Table S2. Specifically, the with-fabrication-error BEST filters designed by PCSED reduces the reconstruction MSE by a factor of ~7.5 when taking those from SED-inv as the reference. The hyperspectral imaging reconstruction comparison is shown in Figure S12-S14. In Both hyperspectral imaging and spectral curves reconstruction, our PCSED method shows a significantly lower MSE.

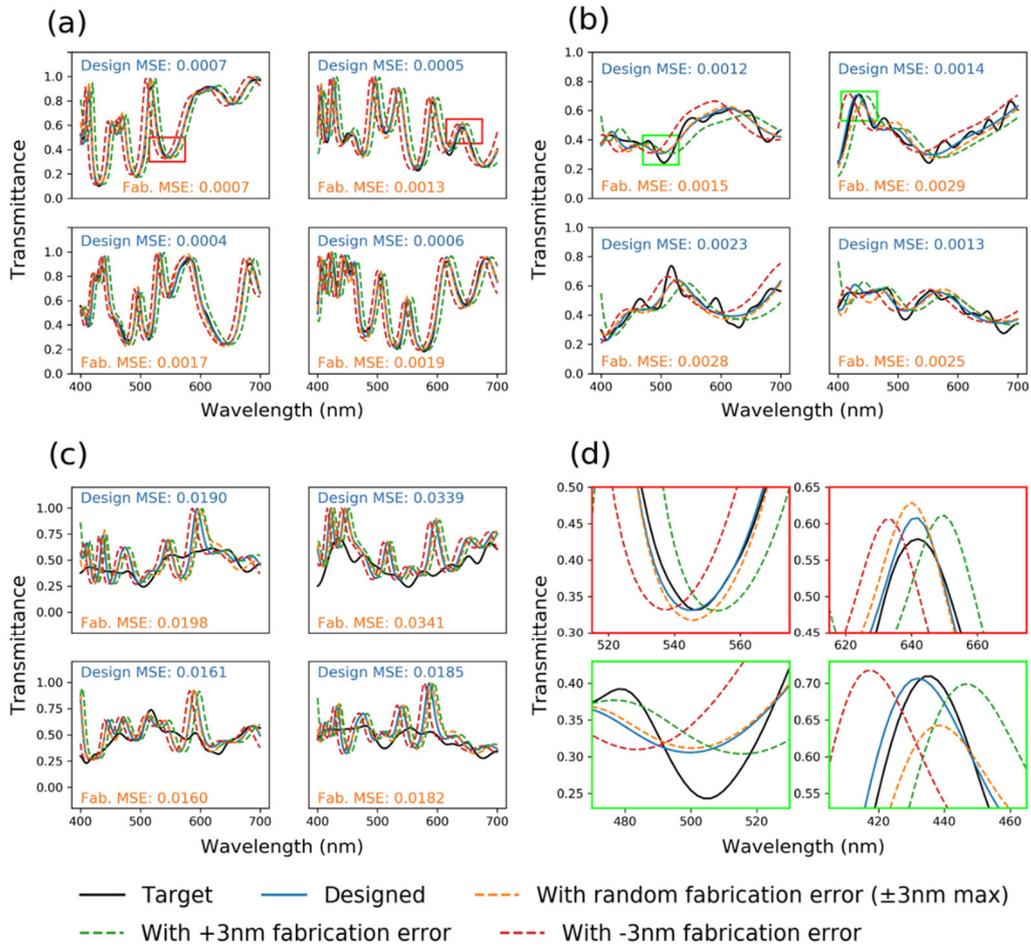

Figure S11. Spectral responses of the target (black solid), designed (blue solid) and synthetically fabricated (dashed) thin-film BEST filters using (a) PCSED method, (b) SED-inv method with a 0–150 nm thickness constraint, and (c) SED-inv method with a 100–300 nm thickness constraint. (d) magnified projected images at the positions highlighted in (a) and (b). The orange dashed curves are responses that simulated by adding ±3 nm maximum random fabrication errors to the designed structure parameters, the green and the red dashed curves are with +3 nm and –3 nm errors, respectively. The 'Design MSE' labeled at the top of each graph represents the MSE between the designed and the target spectral response for each BEST filter and the 'Fab. MSE' labeled at the bottom represents the MSE between the synthetically fabricated and the target spectral response.

**Table S2. Spectral reconstruction MSE of the thin-film spectral camera designed by PCSED and SED-inv.**

|  | PCSED | SED-inv |
|---|---|---|
| Target | 0.00008 | 0.00008 |
| Designed | 0.00046 | 0.00096 |
| W. fab. error | 0.00082 | 0.00602 |

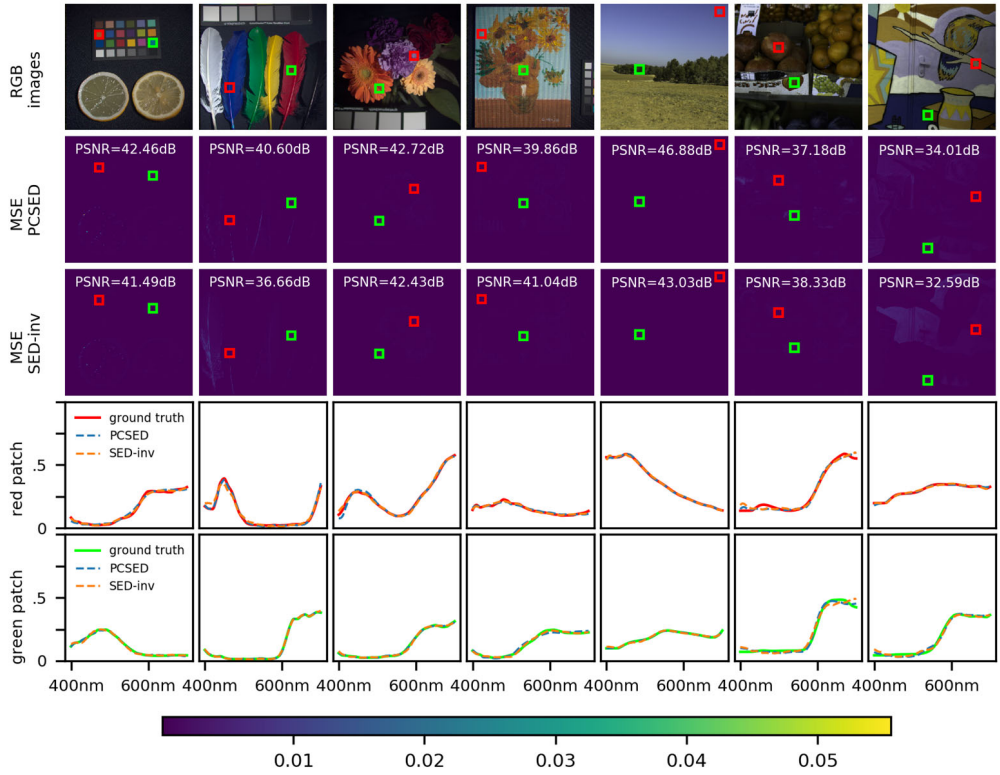

Figure S12. Hyperspectral image reconstruction using the target BEST filters by PCSED and SED-inv method. As there is no design error, both networks reconstruct the hyperspectral images with high accuracy.

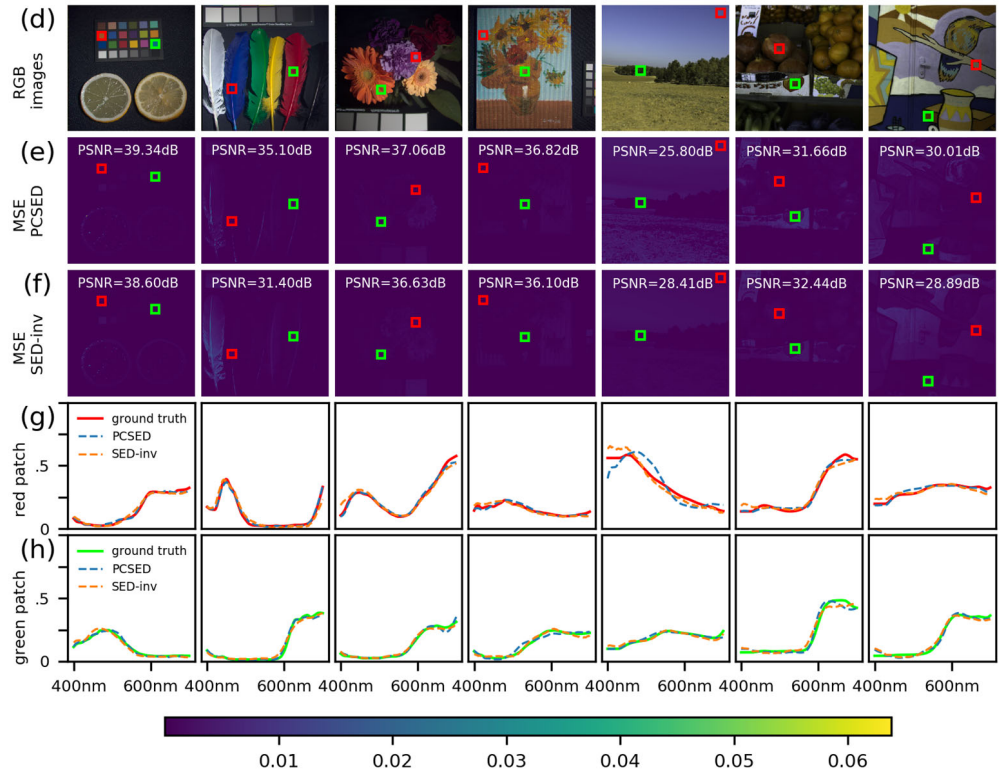

Figure S13. Hyperspectral image reconstruction using the designed BEST filters by PCSED and SED-inv method. With the design error, the SED-inv method shows slightly lower reconstruction accuracy.

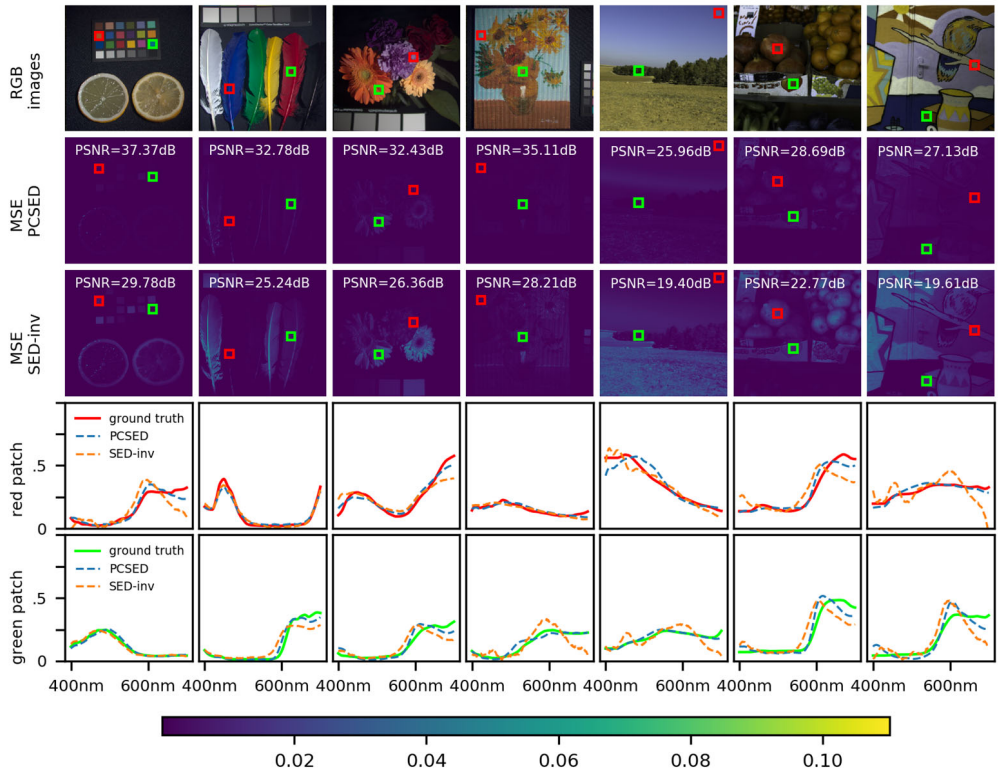

Figure S14. Hyperspectral image reconstruction using the BEST filters with a ±3 nm maximum random fabrication error by PCSED and SED-inv method. With design error and the fabrication error, the PCSED method shows a more stable performance than the SED-inv.